\documentstyle[aps,preprint]{revtex}

\begin{document}
\author{J. Tempere, J. T. Devreese}
\address{Departement Natuurkunde, Universiteit Antwerpen (UIA), Universiteitsplein 1,
B2610 Antwerpen, Belgium. }
\title{Optical Absorption of an Interacting Many-Polaron Gas}
\date{17/11/2000}
\maketitle

\begin{abstract}
The optical absorption of a many (continuum) polaron gas is derived in the
framework of a variational approach at zero temperature and weak or
intermediate electron-phonon coupling strength. We derive a compact formula
for the optical conductivity of the many-polaron system taking into account
many-body effects in the electron or hole system. Within the method
presented here, these effects are contained completely in the dynamical
structure factor of the electron or hole system. This allows to build on
well-established studies of the interacting electron gas. Based on this
approach a novel feature in the absorption spectrum of the many-polaron gas,
related to the emission of a plasmon together with a phonon, is identified.
As an application and illustration of the technique, we compare the
theoretical many-polaron optical absorption spectrum as derived in the
present work with the `d-band' absorption feature in Nd$_{2}$CuO$_{2}$.
Similarities are shown between the theoretically and the experimentally
derived first frequency moment of the optical absorption of a family of
differently doped Nd$_{2-x}$Ce$_{x}$CuO$_{4-y}$ materials.
\end{abstract}


\section{Introduction}

As is the case for polar semiconductors and ionic crystals \cite{ddg},
insight into the nature of polarons in high-temperature superconductors can
be gained by studying the optical properties of these materials. The goal of
the present paper is to present a theory of the optical conductivity of a
system of continuum polarons at any density, including many-body effects
between the constituent charge carriers, for small and intermediate values
of the electron-phonon coupling constant and for zero temperature. The
method we develop here is based on the variational method introduced by
Lemmens, Devreese, Brosens (LDB) \cite{LDB} for the ground state energy of
the many-polaron gas. The advantage of this approach over other theories of
many-polaron optical absorption \cite{catau1,dev1}\ is that it allows to
include the many-body effects in the system of the constituent charge
carriers of the polaron gas on the level of the dynamical structure factor
of the underlying electron (or hole) system. Thus it is possible to select
the level of approximation used in the treatment of the many-polaron gas by
choosing the appropriate expression of the dynamical structure factor for
the {\it electron} (or hole) system.

Recently the infrared spectrum of cuprates has been the subject of intensive
investigations \cite
{calva2,calva0,falck1,calva1,calva1b,crawford,zhang,thomas,homes1},
especially in the case of the neodymium-cerium cuprate family Nd$_{2-x}$Ce$%
_{x}$CuO$_{4-y}$ (NCCO) \cite
{calva2,calva1,calva1b,crawford,zhang,thomas,homes1}. Several optical
absorption features in the infrared cuprate spectrum have been tentatively
associated with large polarons \cite{falck1,jt1} or with a mixture of large
and small (bi)polarons \cite{emin}. These comparisons with polaron theory
were derived using a single-polaron picture, so that the density (doping)
dependence of the optical absorption spectra could not be studied in detail.
The many-body theory of the $N$-polaron spectrum, presented here, allows to
study the density (doping) dependence of optical absorption spectra. As a
first application of the many-polaron optical absorption theory introduced
here, a preliminary comparison is presented between the theoretical
many-polaron optical absorption derived in the current work and the
mid-infrared spectrum of the neodymium-cerium cuprates recently determined
experimentally by Calvani and co-workers\ \cite{calva2}.

\section{Optical absorption in the many-polaron system}

\subsection{The LDB\ variational wave function for a many-polaron system}

The Hamiltonian of a system of $N$ interacting continuum polarons is given
by:

\begin{equation}
H_{0}=\sum_{j=1}^{N}\frac{p_{j}^{2}}{2m_{b}}+\sum_{{\bf k}}\hbar \omega _{%
\text{LO}}a_{{\bf k}}^{+}a_{{\bf k}}+\sum_{{\bf k}}\sum_{j=1}^{N}\left[ e^{i%
{\bf k}.{\bf r}_{j}}a_{{\bf k}}V_{{\bf k}}+e^{-i{\bf k}.{\bf r}_{j}}a_{{\bf k%
}}^{+}V_{{\bf k}}^{\ast }\right] +\frac{e^{2}}{2\varepsilon _{\infty }}%
\sum_{j=1}^{N}\sum_{\ell (\neq j)=1}^{N}\frac{1}{|{\bf r}_{i}-{\bf r}_{j}|},
\label{hmpol}
\end{equation}
where ${\bf r}_{j},{\bf p}_{j}$ represent the position and momentum of the $%
N $ constituent electrons (or holes) with band mass $m_{b}$; $a_{{\bf k}%
}^{+},a_{{\bf k}}$ denote the creation and annihilation operators for
longitudinal optical (LO) phonons with wave vector ${\bf k}$ and frequency $%
\omega _{\text{LO}}$; $V_{{\bf k}}$ describes the amplitude of the
interaction between the electrons and the phonons; and $e$ is the elementary
electron charge. The ground state energy of this many-polaron Hamiltonian
has been studied before by LDB \cite{LDB}, for weak and intermediate
strengths of the electron-phonon coupling, by introducing a variational wave
function: 
\begin{equation}
\left| \psi _{\text{LDB}}\right\rangle =U\left| \phi \right\rangle \left|
\varphi _{\text{el}}\right\rangle ,  \label{psiLDB}
\end{equation}
where $\left| \varphi _{\text{el}}\right\rangle $ represents the
ground-state many-body wave function for the electron (or hole) system and $%
\left| \phi \right\rangle $ is the phonon vacuum, and $U$ is a many-body
unitary operator which determines a canonical transformation for a fermion
gas interacting with a boson field: 
\begin{equation}
U=\exp \left\{ \sum_{j=1}^{N}\sum_{{\bf k}}\left( f_{{\bf k}}a_{{\bf k}}e^{i%
{\bf k}.{\bf r}_{j}}-f_{{\bf k}}^{\ast }a_{{\bf k}}^{+}e^{-i{\bf k}.{\bf r}%
_{j}}\right) \right\} .  \label{U}
\end{equation}
In the limit of one fermion, $U$\ reduces to a canonical transformation
inspired by Tomonaga \cite{tomo} and applied later by several workers, after
Lee, Low and Pines \cite{LLP}, but always for one particle-theories. In LDB 
\cite{LDB}, this canonical transformation was extended and used to establish
a many-fermion theory. The $f_{{\bf k}}$ were determined variationally \cite
{LDB} resulting in 
\begin{equation}
f_{{\bf k}}=%
{\displaystyle{V_{{\bf k}} \over \hbar \omega _{LO}+%
{\displaystyle{\hbar ^{2}k^{2} \over 2m_{b}S({\bf k})}}}}%
,  \label{fk}
\end{equation}
for a system with total momentum ${\bf P=}\sum_{j}{\bf p}_{j}=0$. In this
expression, $S({\bf k})$ represents the static structure factor of the
constituent interacting many electron or hole system : 
\begin{equation}
NS({\bf k})=\left\langle \sum_{j=1}^{N}\sum_{j^{\prime }=1}^{N}e^{i{\bf k}.(%
{\bf r}_{j}-{\bf r}_{j^{\prime }})}\right\rangle .
\end{equation}
The angular brackets $\left\langle \text{...}\right\rangle $\ represent the
expectation value with respect to the ground state. It may be emphasized
that (\ref{U}), although it appears like a straightforward generalization of
the one-particle transformation in \cite{tomo}, represents\ -especially in
its implementation- a nontrivial extension of a one-particle approximation
to a many-body system. As noted in the introduction, the main advantage of
the LDB many-polaron variational approach lies in the fact that the
many-body effects in the system of charge carriers (electrons or holes) are
completely contained in the structure factor of the electron (or hole) gas.
This advantage will be carried through into the calculation of the optical
properties of the interacting gas of continuum polarons which is the subject
of the current paper.

\subsection{Kubo formula for the optical conductivity of the many-polaron gas%
}

The many-polaron optical conductivity is the response of the
current-density, in the system described by the Hamiltonian (\ref{hmpol}),
to an applied electric field (along the $x$-axis) with frequency $\omega $.
This applied electric field introduces a perturbation term in the
Hamiltonian (\ref{hmpol}), which couples the vector potential of the
incident electromagnetic field to the current-density. As is well known,
within linear response theory, the optical conductivity can be expressed
through the Kubo formula as a current-current correlation function \cite
{Mahan}: 
\begin{equation}
\sigma (\omega )=i\frac{Ne^{2}}{\text{V}m_{b}\omega }+\frac{1}{\text{V}\hbar
\omega }\int_{0}^{\infty }e^{i\omega t}\left\langle \left[ J_{x}(t),J_{x}(0)%
\right] \right\rangle dt.
\end{equation}
In this expression, V is the volume of the system, and $J_{x}$ is the $x$%
-component of the current operator ${\bf J},$ which is related to the
momentum operators of the charge carriers: 
\begin{equation}
{\bf J}=\frac{q}{m_{b}}\sum_{j=1}^{N}{\bf p}_{j}=\frac{q}{m_{b}}{\bf P,}
\end{equation}
with $q$ the charge of the charge carriers ($+e$ for holes, $-e$ for
electrons) and ${\bf P}$ the total momentum operator of the charge carriers.
The real part of the optical conductivity at temperature zero, which is
proportional to the optical absorption coefficient, can be written as a
function of the total momentum operator of the charge carriers as follows : 
\begin{equation}
\mathop{\rm Re}%
[\sigma (\omega )]=\frac{1}{\text{V}\hbar \omega }\frac{e^{2}}{m_{b}^{2}}%
\mathop{\rm Re}%
\left\{ \int_{0}^{\infty }e^{i\omega t}\left\langle \left[ P_{x}(t),P_{x}(0)%
\right] \right\rangle dt\right\} .
\end{equation}
Integrating by parts twice, the real part of the optical conductivity of the
many-polaron system can be written with a force-force correlation function: 
\begin{equation}
\mathop{\rm Re}%
[\sigma (\omega )]=\frac{1}{\text{V}\hbar \omega ^{3}}\frac{e^{2}}{m_{b}^{2}}%
\mathop{\rm Re}%
\left\{ \int_{0}^{\infty }e^{i\omega t}\left\langle \left[ F_{x}(t),F_{x}(0)%
\right] \right\rangle dt\right\} ,
\end{equation}
with ${\bf F}=(i/\hbar )\left[ H_{0},{\bf P}\right] $. The commutator of the
Hamiltonian (\ref{hmpol}) with the total momentum operator of the charge
carriers simplifies to 
\begin{equation}
{\bf F}=-i\sum_{{\bf k}}\sum_{j=1}^{N}{\bf k}\left( e^{i{\bf k}.{\bf r}%
_{j}}a_{{\bf k}}V_{{\bf k}}+e^{-i{\bf k}.{\bf r}_{j}}a_{{\bf k}}^{+}V_{{\bf k%
}}^{\ast }\right) .
\end{equation}
This result for the force operator clarifies the significance of using the
force-force correlation function rather than the momentum-momentum
correlation function. The operator product $F_{x}(t)F_{x}(0)$\ is
proportional to $|V_{{\bf k}}|^{2}$, the charge carrier - phonon interaction
strength. This will be a distinct advantage for any expansion of the final
result in the charge carrier - phonon interaction strength, since one power
of $|V_{{\bf k}}|^{2}$\ is factored out beforehand. Denoting $\rho _{{\bf k}%
}=\sum_{j=1}^{N}e^{i{\bf k}.{\bf r}_{j}}$, the real part of the optical
conductivity takes the form: 
\begin{equation}
\mathop{\rm Re}%
[\sigma (\omega )]=\frac{1}{\text{V}\hbar \omega ^{3}}\frac{e^{2}}{m_{b}^{2}}%
\mathop{\rm Re}%
\left\{ \sum_{{\bf k,k}^{\prime }}k_{x}.k_{x}^{\prime }\int_{0}^{\infty
}e^{i\omega t}\left\langle \left[ 
\begin{array}{c}
e^{iH_{0}t/\hbar }\left( \rho _{{\bf k}}a_{{\bf k}}V_{{\bf k}}+\rho _{-{\bf k%
}}a_{{\bf k}}^{+}V_{{\bf k}}^{\ast }\right) e^{-iH_{0}t/\hbar }, \\ 
\left( \rho _{{\bf k}^{\prime }}a_{{\bf k}^{\prime }}V_{{\bf k}^{\prime
}}+\rho _{-{\bf k}^{\prime }}a_{{\bf k}^{\prime }}^{+}V_{{\bf k}^{\prime
}}^{\ast }\right)
\end{array}
\right] \right\rangle _{0}dt\right\} .  \label{resig0}
\end{equation}
Up to this point, no approximations other than {\it linear response theory}
have been made.

\subsection{LDB\ canonical transformation for the optical conductivity}

The expectation value appearing in the right hand side of expression (\ref
{resig0}) for the real part of the optical conductivity is calculated now
with respect to the LDB many-polaron wave function $\left| \psi _{\text{LDB}%
}\right\rangle $ (\ref{psiLDB}): 
\begin{eqnarray}
{\cal J}({\bf k},{\bf k}^{\prime }) &=&\left\langle \psi _{\text{LDB}}\left| %
\left[ e^{iH_{0}t/\hbar }\left( \rho _{{\bf k}}a_{{\bf k}}V_{{\bf k}}+\rho
_{-{\bf k}}a_{{\bf k}}^{+}V_{{\bf k}}^{\ast }\right) e^{-iH_{0}t/\hbar
},\left( \rho _{{\bf k}^{\prime }}a_{{\bf k}^{\prime }}V_{{\bf k}^{\prime
}}+\rho _{-{\bf k}^{\prime }}a_{{\bf k}^{\prime }}^{+}V_{{\bf k}^{\prime
}}^{\ast }\right) \right] \right| \psi _{\text{LDB}}\right\rangle \\
&=&\left\langle \varphi _{\text{el}}\left| \left\langle \phi \left| \left[ 
\begin{array}{c}
e^{iH^{\prime }t/\hbar }U^{-1}\left( \rho _{{\bf k}}a_{{\bf k}}V_{{\bf k}%
}+\rho _{-{\bf k}}a_{{\bf k}}^{+}V_{{\bf k}}^{\ast }\right) Ue^{-iH^{\prime
}t/\hbar }, \\ 
U^{-1}\left( \rho _{{\bf k}^{\prime }}a_{{\bf k}^{\prime }}V_{{\bf k}%
^{\prime }}+\rho _{-{\bf k}^{\prime }}a_{{\bf k}^{\prime }}^{+}V_{{\bf k}%
^{\prime }}^{\ast }\right) U
\end{array}
\right] \right| \phi \right\rangle \right| \varphi _{\text{el}}\right\rangle
,
\end{eqnarray}
where $U$ is the many-polaron canonical transformation defined in (\ref{U})
and $H^{\prime }=U^{-1}H_{0}U$ is the transformed Hamiltonian obtained in 
\cite{LDB}. The canonical transformation of the force term is 
\begin{equation}
U^{-1}\left( \rho _{{\bf k}}a_{{\bf k}}V_{{\bf k}}+\rho _{-{\bf k}}a_{{\bf k}%
}^{+}V_{{\bf k}}^{\ast }\right) U=\rho _{{\bf k}}\left( a_{{\bf k}}-f_{{\bf k%
}}^{\ast }\rho _{-{\bf k}}\right) V_{{\bf k}}+\rho _{-{\bf k}}\left( a_{{\bf %
k}}^{+}-f_{{\bf k}}\rho _{{\bf k}}\right) V_{{\bf k}}^{\ast }.
\end{equation}
The terms of lowest order in the electron-phonon interaction amplitude $|V_{%
{\bf k}}|^{2}$ are given by 
\begin{eqnarray}
{\cal J}({\bf k},{\bf k}^{\prime }) &=&|V_{{\bf k}}|^{2}\delta _{{\bf kk}%
^{\prime }}\left\langle \varphi _{\text{el}}\left| \left\langle \phi \left|
e^{iH^{\prime }t/\hbar }\rho _{{\bf k}}a_{{\bf k}}e^{-iH^{\prime }t/\hbar
}\rho _{-{\bf k}}a_{{\bf k}}^{+}-\rho _{{\bf k}}a_{{\bf k}}e^{iH^{\prime
}t/\hbar }\rho _{-{\bf k}}a_{{\bf k}}^{+}e^{-iH^{\prime }t/\hbar }\right|
\phi \right\rangle \right| \varphi _{\text{el}}\right\rangle \\
&=&2i|V_{{\bf k}}|^{2}\delta _{{\bf kk}^{\prime }}%
\mathop{\rm Im}%
\left[ \left\langle \varphi _{\text{el}}\left| \left\langle \phi \left|
e^{iH^{\prime }t/\hbar }\rho _{{\bf k}}a_{{\bf k}}e^{-iH^{\prime }t/\hbar
}\rho _{-{\bf k}}a_{{\bf k}}^{+}\right| \phi \right\rangle \right| \varphi _{%
\text{el}}\right\rangle \right] .
\end{eqnarray}
Taking the expectation value with respect to the phonon vacuum, we find 
\begin{equation}
{\cal J}({\bf k},{\bf k}^{\prime })=2i|V_{{\bf k}}|^{2}\delta _{{\bf kk}%
^{\prime }}%
\mathop{\rm Im}%
\left\{ e^{-i\omega _{\text{LO}}t}\left\langle \varphi _{\text{el}}\left|
e^{iH^{\prime }t/\hbar }\rho _{{\bf k}}e^{-iH^{\prime }t/\hbar }\rho _{-{\bf %
k}}\right| \varphi _{\text{el}}\right\rangle \right\} .
\end{equation}
This can be substituted in the expression (\ref{resig0}) for the real part
of the optical conductivity, which becomes 
\begin{equation}
\mathop{\rm Re}%
[\sigma (\omega )]=-2\frac{1}{\text{V}\hbar \omega ^{3}}\frac{e^{2}}{%
m_{b}^{2}}%
\mathop{\rm Im}%
\left\{ \sum_{{\bf k}}k_{x}^{2}|V_{{\bf k}}|^{2}\int_{0}^{\infty }e^{i\omega
t}%
\mathop{\rm Im}%
\left[ e^{-i\omega _{\text{LO}}t}\left\langle \varphi _{\text{el}}\left|
e^{iH^{\prime }t/\hbar }\rho _{{\bf k}}e^{-iH^{\prime }t/\hbar }\rho _{-{\bf %
k}}\right| \varphi _{\text{el}}\right\rangle \right] dt\right\} .
\label{resig1}
\end{equation}
The right hand side of (\ref{resig1}) can be written in a more compact form
by introducing the dynamical structure factor\ of the electron (or hole)
system.

\subsection{General expression}

To find the formula for the real part of the optical conductivity in its
final form, we introduce the standard expression for the dynamical structure
factor of the system of charge carriers interacting through a Coulomb
potential,

\begin{equation}
S({\bf q},w)=%
\displaystyle\int %
\limits_{-\infty }^{+\infty }\left\langle \varphi _{\text{el}}\left| 
{\displaystyle{1 \over 2}}%
\mathop{\displaystyle\sum}%
\limits_{j,\ell }e^{i{\bf q}.({\bf r}_{j}(t)-{\bf r}_{\ell }(0))}\right|
\varphi _{\text{el}}\right\rangle e^{iwt}dt.
\end{equation}
Rewriting expression (\ref{resig1}) with the dynamical structure factor of
the electron (or hole) gas results in: 
\begin{mathletters}
\begin{equation}
\mathop{\rm Re}%
[\sigma (\omega )]=\frac{n}{\hbar \omega ^{3}}\frac{e^{2}}{m_{b}^{2}}\sum_{%
{\bf k}}k_{x}^{2}|V_{{\bf k}}|^{2}S({\bf k},\omega -\omega _{\text{LO}}), 
\eqnum{20}  \label{opticabs}
\end{equation}
where $n=N/$V is the density of charge carriers. As noted before, $V_{{\bf k}%
}$ is the electron-phonon interaction amplitude and $k_{x}$ is the $x$%
-component of the wave vector. Formula (\ref{opticabs}) for the optical
absorption of the many-polaron system has an intuitively appealing form.

In the theory of one-polaron optical absorption for weak coupling constants $%
\alpha $, the optical absorption coefficient as obtained from Fermi's golden
rule is \cite{Greenbook2}: 
\end{mathletters}
\begin{equation}
\text{1 polaron}:%
\mathop{\rm Re}%
[\sigma (\omega )]\propto \omega ^{-3}\sum_{{\bf k}}k_{x}^{2}|V_{{\bf k}%
}|^{2}\delta \lbrack \hbar k^{2}/(2m_{\text{b}})-(\omega -\omega _{\text{LO}%
})].  \label{wanpol}
\end{equation}
At low densities, the dynamical structure factor $S(q,\nu )$\ is strongly
peaked around $q^{2}/2=\nu $\ and is close to zero everywhere else \cite
{Mahan2}. Substituting a delta-function $\delta (q^{2}/2-\nu )$\ for the
dynamical structure factor in formula (\ref{opticabs}) it is easily seen
that the one-polaron limit (\ref{wanpol}) \cite{Greenbook2} is retrieved.
The one-polaron result (\ref{wanpol}) is derived \cite{Greenbook2} by
considering a process in which the initial state consists of a photon of
energy $\hbar \omega $ and a polaron in its ground state, and the final
state consists of an emitted LO phonon with energy $\hbar \omega _{\text{LO}}
$\ and the polaron, scattered into a state with momentum ${\bf k}$\ and
kinetic energy $(\hbar k)^{2}/(2m_{\text{b}})=\hbar (\omega -\omega _{\text{%
LO}}).$\ The many-polaron result, formula (\ref{opticabs}), is a
generalization of this one-polaron picture. The contribution which
corresponds to the scattering of a polaron into the momentum state ${\bf k}$
and energy $\hbar (\omega -\omega _{\text{LO}})$ is now weighed by the
dynamical structure factor $S(k,\omega -\omega _{\text{LO}})$ of the
electron (or hole) gas.

Formula (\ref{opticabs}) is reminiscent of the Hopfield formula \cite
{hopfield} describing the effect of impurities on the optical absorption of
metals, which in turn is related to the expression obtained by Ron and Tzoar 
\cite{RT} for the optical absorption in a quantum plasma. Formula (\ref
{opticabs}) also represents a generalization of the results obtained by
Gurevich, Lang and Firsov \cite{gur}. These authors focused their attention
on the many-body effects related to the Fermi exclusion statistics, whereas
the present analysis will extend the results of \cite{gur} to study the
influence of plasmons and further many-body effects in the system of the
constituent electrons or holes, as discussed in the next section.

The advantage of the LDB canonical transformation method \cite{LDB} for the
evaluation of {\it the ground state energy }of a polaron gas is that the
many-body effects are contained in the {\it static }structure factor of the
electron (or hole) system, appearing in the analytical expression for the
energy. The corresponding advantage of the canonical transformation in the
present case, for {\it the optical conductivity}, is that many-body effects
are again incorporated through a structure factor, now the {\it dynamical }%
structure factor of the electron (or hole) system. The level of
approximation made in treating the many-body nature of the polaron system is
determined by the choice of the dynamical structure factor of the electron
or hole system.

In the present treatment of the {\it electron-phonon }interactions, the
terms to leading order in $|V_{{\bf k}}|^{2}$ are automatically taken into
account through the variational formulation based on LDB \cite{LDB}. As
noted before, the use of the force-force correlation function allows to
express the real part of the optical conductivity as $%
\mathop{\rm Re}%
[\sigma (\omega )]=\alpha {\cal F}(\omega ,\alpha ),$ so that to lowest
order in $\alpha $, $%
\mathop{\rm Re}%
[\sigma (\omega )]=\alpha {\cal F}(\omega ,\alpha =0)$ where the
electron-phonon interaction is no longer present in the factor ${\cal F}$\
which includes the many-body effects of the electron (or hole) system. A
possible way to take into account higher-order terms in the electron-phonon
interactions would be to include electron-phonon coupling effects at the
level of the dynamical structure factor appearing in ${\cal F}$, in a manner
similar to Mahan's treatment of the polaron spectral function \cite{mah2},
or to include multiple-phonon final states in the calculation \cite
{HuybrechtsPRB8}.

\subsection{Scaling relation for the optical absorption in two and three
dimensions}

The modulus squared of the Fr\"{o}hlich electron-phonon interaction
amplitude is given by 
\begin{equation}
|V_{{\bf k}}|^{2}=\left\{ 
\begin{array}{l}
{\displaystyle{(\hbar \omega _{\text{LO}})^{2} \over k^{2}}}%
{\displaystyle{4\pi \alpha  \over \text{V}}}%
\sqrt{%
{\displaystyle{\hbar  \over 2m_{b}\omega _{\text{LO}}}}%
}\text{ in 3D} \\ 
{\displaystyle{(\hbar \omega _{\text{LO}})^{2} \over k}}%
{\displaystyle{2\pi \alpha  \over \text{A}}}%
\sqrt{%
{\displaystyle{\hbar  \over 2m_{b}\omega _{\text{LO}}}}%
}\,\text{in 2D,}
\end{array}
\right.
\end{equation}
where $\alpha $ is the (dimensionless) Fr\"{o}hlich coupling constant
determining the coupling strength between the charge carriers and the
longitudinal optical phonons, and A is the surface of the 2D system \cite
{scaling}. In what follows, we will use polaron units ($\hbar =m_{b}=\omega
_{\text{LO}}=1$). The sum over wave vectors in (\ref{opticabs}) can be
written as an integral, so that for the three dimensional case (with
dynamical structure factor $S_{\text{3D}}$) we find: 
\begin{equation}
\mathop{\rm Re}%
[\sigma _{\text{3D}}(\omega )]=ne^{2}\text{ }\frac{2}{3}\alpha \frac{1}{2\pi
\omega ^{3}}%
\displaystyle\int %
\limits_{0}^{\infty }dq\text{ }q^{2}S_{\text{3D}}(q,\omega -\omega _{\text{LO%
}}),  \label{sig3d}
\end{equation}
and for the two-dimensional case: 
\begin{equation}
\mathop{\rm Re}%
[\sigma _{\text{2D}}(\omega )]=ne^{2}\text{ }\frac{\pi }{2}\alpha \frac{1}{%
2\pi \omega ^{3}}%
\displaystyle\int %
\limits_{0}^{\infty }dq\text{ }q^{2}S_{2\text{D}}(q,\omega -\omega _{\text{LO%
}}).  \label{sig2d}
\end{equation}
From these expressions, it is clear that the scaling relation 
\begin{equation}
\mathop{\rm Re}%
[\sigma _{\text{2D}}(\omega ,\alpha )]=%
\mathop{\rm Re}%
[\sigma _{\text{3D}}(\omega ,3\pi \alpha /4)]
\end{equation}
which holds for the one-polaron case introduced in ref. \cite{scaling}, is
also valid for the many-polaron case if the corresponding 2D or 3D dynamical
structure factor is used.

\section{Results and discussion}

\subsection{General results}

The expressions (\ref{sig3d},\ref{sig2d}) allow us to derive results both
for a three-dimensional and for a two dimensional polaron gas at $T=0$. The
choice of a dynamical structure factor for the electron (or hole) system
allows furthermore to study the different levels of approximation
(Hartree-Fock, RPA,...) in the treatment of the many-electron or many-hole
system. The results presented in this section were obtained using the
material parameters of GaAs (for the two-dimensional case) and ZnO (for the
three-dimensional case). These material parameters \cite
{calva1,wu2,GreenBook,AlonsoPRB55} are summarized in Table I.

Fig. 1 shows the Hartree-Fock and the RPA result for the 2D many-polaron
optical absorption spectrum (for GaAs, at a density $n=10^{12}$ cm$^{-2}$).
For reference, the dashed curve represents the familiar one-polaron result.
In a first step, we discuss the result obtained by using the Hartree-Fock
expression for the dynamical structure factor of the electron (or hole)
system in the expressions (\ref{sig3d},\ref{sig2d}).

The Fermi statistics cause the polarons to fill up a Fermi sphere up to $k_{%
\text{F}}=[n/(2\pi )]^{1/2}$. The optical absorption of the polaron gas
resulting from this system is represented by the full curve labeled
`Hartree-Fock' in Fig. 1. The spectral weight at frequencies between $\omega
_{\text{LO}}$ and $1.4$\ $\omega _{\text{LO}}$\ in Fig. 1 is reduced as
compared to the single polaron case, whereas at higher frequencies it is
enhanced. A kink appears in the spectrum at $\omega =\omega _{\text{LO}}+E_{%
\text{F}}/\hbar $, as indicated by the dotted vertical line in Fig. 1.

This can be understood as follows. The absorption process is characterized
by an {\it initial }state consisting of a polaron gas filling up the Fermi
sphere (up to energy $E_{\text{F}}$, at $T=0$) and a photon with given
energy $\hbar \omega $, and by a {\it final} state made up of an emitted LO
phonon with energy $\hbar \omega _{\text{LO}}$ and a polaron gas such that
one polaron state inside the Fermi sphere is not occupied and one polaron
state with energy $E>E_{\text{F}}$ is occupied. The incident photon can only
excite polarons out of the Fermi sea for which $\hbar \omega >h\omega _{%
\text{LO}}+E_{\text{F}}$. A straightforward calculation in 2D shows that the
fraction of polaron states in the Fermi sphere which can interact with a
photon of energy $\hbar \omega $\ is given by 
\begin{equation}
\left\{ 
\begin{array}{l}
0\text{ for }\omega <\omega _{\text{LO}} \\ 
{\displaystyle{\hbar (\omega -\omega _{\text{LO}}) \over E_{\text{F}}}}%
\text{ for }\omega _{\text{LO}}<\omega <E_{\text{F}}/\hbar +\omega _{\text{LO%
}} \\ 
1\text{ for }\omega >E_{\text{F}}/\hbar +\omega _{\text{LO}}
\end{array}
.\right.  \label{frpart}
\end{equation}
For photon frequencies between $\omega _{\text{LO}}$ and $\omega _{\text{LO}%
}+E_{\text{F}}/\hbar $, the number of polarons which cannot participate in
the optical absorption process due to the Pauli exclusion principle,
decreases linearly. At $\omega =\omega _{\text{LO}}+E_{\text{F}}/\hbar $,
all polarons can participate. This leads to a kink in the function (\ref
{frpart}) describing the number of polarons which can interact with the
photon of given energy $\hbar \omega $, as a function of $\omega $. This is
also the origin of the kink in the optical absorption. This kink in the 2D
many-polaron optical absorption spectrum at $\hbar \omega =\hbar \omega _{%
\text{LO}}+E_{\text{F}}$\ was already noted in \cite{wu2}.

The full curve labeled with `RPA' in Fig. 1 is obtained by using the random
phase approximation (RPA) for the dynamical structure factor of the electron
(or hole) system. It illustrates the combined effects of the Fermi
statistics, discussed in the previous paragraph, and screening in the
electron (or hole) system. In comparison to the Hartree-Fock curve, the main
effect is an overall reduction of the spectral weight at frequencies $\omega
>\omega _{\text{LO}}${\bf . }There is however a second effect, which is the
appearance of a contribution related to plasmons - this is the subject of
the next subsection.

\subsection{Plasmon-phonon contribution}

The RPA dynamical structure factor for the electron (or hole) system can be
separated in two parts, one related to continuum excitations of the
electrons (or holes) $S_{\text{cont}}$\ , and one related to the undamped
plasmon branch \cite{isihara}: 
\[
S_{\text{RPA}}(q,\omega )=A_{\text{pl}}(q)\delta \lbrack \omega -\omega _{%
\text{pl}}(q)]+S_{\text{cont}}(q,\omega ), 
\]
where $\omega _{\text{pl}}(q)$\ is the wave number dependent plasmon
frequency and $A_{\text{pl}}$\ is the strength of the undamped plasmon
branch \cite{isihara}. The insets of Fig. 2 depict the regions in the $q$-$v$
plane ($q=k/k_{F},v=m_{\text{b}}\omega /(\hbar k_{F}^{2})$)\ where the RPA
dynamical structure factor is different from zero. The contribution (after
substitution of $S_{\text{RPA}}$\ in (\ref{sig3d})-(\ref{sig2d})) in the
many-polaron optical absorption) deriving from the undamped plasmon branch $%
A_{\text{pl}}(q)\delta \lbrack \omega -\omega _{\text{pl}}(q)]$\ will be
denoted as {\it `plasmon-phonon' contribution}. The physical process related
to this contribution is the emission of both a phonon and a plasmon in the
scattering process.

Figure 2 shows the result for the optical absorption of the many-polaron gas
for the 2D case\ (GaAs, left panel) and the 3D case (ZnO, right panel). For
reference, the dashed curves show the one-polaron result. The full curves
show the many-polaron results in the random-phase approximation. The shaded
gray areas indicate the plasmon-phonon contribution.

Now examine the 3D case (the right panel of Fig. 2). The frequency of the
undamped plasmon mode lies between $\omega _{1}$ and $\omega _{\text{2}}$
where $\omega _{\text{1}}=\omega _{\text{pl}}=\sqrt{4\pi ne^{2}/m_{\text{b}}}
$ is the frequency of the plasmon branch at $q=0$, and $\omega _{2}$ is the
frequency at which the branch of the undamped plasmons enters the Landau
damping region (whose edge is given by $\omega =\hbar q^{2}/(2m_{b})+\hbar
k_{\text{F}}q/(2m_{b})$). The corresponding plasmon-phonon contribution to
the optical absorption `starts' at $\omega _{\text{LO}}+\omega _{1}$ and
`ends' at $\omega _{\text{LO}}+\omega _{2}$. These frequencies are indicated
by vertical dotted lines in the right panel of Fig. 2.

In the 2D case, the undamped plasmon branch is acoustic-like; for $%
q\rightarrow 0$, $\omega _{\text{pl}}\rightarrow 0$. Consequently, the
phonon-plasmon peak in this case extends from $\omega _{\text{LO}}$ up to $%
\omega _{\text{LO}}+\omega _{2}$ where $\omega _{2}$ is the frequency at
which the undamped plasmon branch enters the region of the continuum
excitations of the 2D (RPA) electron gas.

In Fig. 3, the evolution of the many-polaron optical absorption spectrum is
shown as the density of electrons (or holes) is increased. Two effects can
be observed for increasing density: the reduction of the optical absorption
above $\omega >\omega _{\text{LO}}${\bf \ }and the shift towards higher
frequencies of the plasmon-phonon contribution, both in 2D (left panel) and
in 3D (right panel). The f-sum rule is nevertheless satisfied due to the
presence of a central $\delta (\omega )$ peak \cite{DevreesePRB15} in the
optical absorption of the polaron gas at $T=0$.

\subsection{Comparison with other theories}

In earlier work, Wu, Peeters and Devreese \cite{wu2} studied the influence
of screening on the electron-phonon interaction in a two-dimensional
electron gas based on a memory function approach using a perturbation
expansion in the electron-LO phonon coupling constant. The results of this
perturbative approach of \cite{wu2} for the optical conductivity in 2D, also
in the RPA framework, are consistent with the results derived from the
present method based on the {\it variational} LDB unitary transformation.
These authors found an enhancement of the optical absorption at the
frequency where the undamped plasmon branch reaches the region of continuum
excitations of the electron gas. The present, variational, method extends
these results by taking into account the entire undamped plasmon branch.

Recently, Cataudella, De Filippis and Iadonisi \cite{catau1} investigated
the optical properties of the many-polaron gas by calculating the correction
due to electron-phonon interactions to the RPA dielectric function of the
electron gas, starting from the Feynman polaron model and ref. \cite{ddg}.
An aspect of the present method is that it is not restricted to the
random-phase approximation for the treatment of the many-body effects
between the charge carriers. Cataudella {\it et al.} \cite{catau1} also find
a suppression of the optical absorption with increasing density. To our
knowledge, the plasmon-phonon contribution was not revealed by the work of
Cataudella {\it et al. }\cite{catau1}.

\subsection{Comparison to the infrared spectrum of Nd$_{2-x}$Ce$_{x}$CuO$%
_{2-y}$}

Calvani and collaborators have performed doping-dependent measurements of
the infrared absorption spectra of the high-T$_{c}$ material Nd$_{2-x}$Ce$%
_{x}$CuO$_{2-y}$ (NCCO). The region of the spectrum examined by these
authors (50-10000 cm$^{-1}$) is very rich in absorption features: they
observe is a ``Drude-like'' component at the lowest frequencies, and a set
of sharp absorption peaks related to phonons and infrared active modes
(IRAV, up to about 1000 cm$^{-1}$) possibly associated to small (Holstein)
polarons \cite{Fratini}. Three distinct absorption bands can be
distinguished: the `d-band' (around 1000 cm$^{-1}$), the Mid-Infrared band
(MIR, around 5000 cm$^{-1}$) and the Charge-Transfer band (CT, around 10$%
^{4} $ cm$^{-1}$) \cite{calva1}. Of all these features, the d-band and, at a
higher temperatures, the Drude-like component have (hypothetically) been
associated with large polaron optical absorption \cite{calva2,falck1,jt1}.

For the lowest levels of Ce doping, the d-band can be most clearly
distinguished from the other features. The experimental optical absorption
spectrum (up to 3000 cm$^{-1}$) of Nd$_{2}$CuO$_{2-\delta }$ ($\delta <0.004$%
), obtained by Calvani and co-workers \cite{calva2}, is shown in Fig. 4
(shaded area) together with the theoretical curve obtained by the present
method (full, bold curve) and, for reference, the one-polaron optical
absorption result (dotted curve). At lower frequencies (600-1000 cm$^{-1}$)
a marked difference between the single polaron optical absorption and the
many-polaron result is manifest. The experimental d-band can be clearly
identified, rising in intensity at about 600 cm$^{-1}$, peaking around 1000
cm$^{-1}$, and then decreasing in intensity above that frequency. At a
density of $n=1.5$ $10^{17}$ cm$^{-3}$, we found a remarkable agreement
between our theoretical predictions and the experimental curve. The
experimentally determined material parameters used in the present
calculation are summarized in Table I. A background contribution, taken to
be constant over the frequency range of the d-band, was substracted in
Figure 4. The lack of experimental data on several material constants leaves
us with three adjustable parameters: the electron-phonon coupling constant $%
\alpha ,$ the band mass $m_{\text{b}}$, and the density of charge carriers.
These parameters were chosen as follows:

\begin{itemize}
\item  A set of theoretical optical absorption spectra were generated for
different values of the band mass and densities (for $m_{\text{b}%
}=0.1,0.2,0.5,0.8,1.0,2.0$\ and $n=\{0.1,0.2,0.5,1.0,1.2,1.5,2.0\}\times
10^{17}$\ cm$^{-3}$).

\item  For each of those spectra, the coupling constant $\alpha $ was chosen
so as to fit the tail region of the experimental optical absorption spectrum
best, using a least squares fitting procedure (the tail region is relatively
insensitive to many-polaron effects).

\item  The best fitting curve, using again a least squares evaluation of the
goodness of fit, was selected; we found fair agreement with $m_{\text{b}%
}=0.5 $ $m_{\text{e}}$ and $n=1.5$ $10^{17}$\ cm$^{-3}$\ at $\alpha =2.1.$
\end{itemize}

This comparison with experiment could not be performed at higher doping
content:\ the frequency region of the d-band usually contains a strong
non-uniform contribution of other optical absorption features (such as the
onset of the MIR, the tail of the Drude contribution, and the IRAV and
phonon modes). To take these other contributions into account, additional
adjustable parameters would have to be introduced making a comparison less
convincing. Fortunately, experimental results are available in the form of
the normalized first frequency moment of the optical absorption spectrum
(after substraction of MIR$\;$and CT band) $%
\mathop{\rm Re}%
[\sigma _{\text{exp}}(\omega )]$ : 
\begin{equation}
\left\langle \omega \right\rangle =%
{\displaystyle{%
\displaystyle\int \limits_{0}^{\omega _{\text{max}}}\omega %
\mathop{\rm Re}[\sigma _{\text{exp}}(\omega )]d\omega  \over %
\displaystyle\int \limits_{0}^{\omega _{\text{max}}}%
\mathop{\rm Re}[\sigma _{\text{exp}}(\omega )]d\omega }}%
\end{equation}
where $\omega _{\text{max}}=10000$ cm$^{-1}$ \cite{calva2}. Calvani and
co-workers \cite{calva2} determined $\left\langle \omega \right\rangle $ for
NCCO samples with a varying cerium doping content. Increasing the cerium
doping will inject electrons in the copper-oxide planes of the material, and
increase the 2D charge carrier density in these planes. A comparison of this
experimental normalized first frequency moment to the theoretical one
presents the advantage that fewer parameters need to be adapted: only the 
{\it density} and the electron {\it band mass }have to be taken from
experiment or (if experimental values are lacking) fitted.

The carrier density can be estimated numerically from the effective carrier
concentrations in the different samples \cite{calva2} and from a measurement
of the two-dimensional Fermi velocity performed for one of the samples \cite
{homes1}. As for the other cuprates, the band mass of the electrons in NCCO
has not yet been determined experimentally \cite{massa}, and remains as an
adjustable parameter.

Fig. 5 represents the comparison between the present theory and experiment.
The squares with error bars show the experimental results for differently
doped samples of$_{{}}$ NCCO, reported in \cite{calva2}. The dashed curve
shows the normalized first frequency moment of the theoretical optical
absorption spectrum, integrated over the entire frequency range ($\omega _{%
\text{max}}\rightarrow \infty $). The tail region of the many-polaron
optical absorption still carries a significant weight, just as it does in
the one-polaron optical absorption. It is necessary to include the cutoff
frequency. The full curve represents the theoretical first frequency moment
with a cutoff frequency $\omega _{\text{max}}=10000$ cm$^{-1}$, which
corresponds to the experimental cutoff \cite{calva2}.

There exists a fair agreement between the theoretical and the experimental
values of the normalized first frequency moment for the five samples with
lowest density, which have a cerium doping content of $x<0.12$. These
correspond to the squares to the left of the dotted vertical line in Fig. 5.
For the four remaining samples $(x>0.12)$ a discrepancy between the
theoretically predicted first frequency moment for unpaired polarons and the
observed first frequency moment appears. It has been observed experimentally
that the weight of the low-frequency component in these samples (with $%
x>0.12 $) is significantly larger that the corresponding weight in samples
with $x<0.12$ \cite{calva1}. This was interpreted in \cite{calva1} as a
consequence of an {\it insulator-to-metal} transition taking place around
the cerium-doping level of $x=0.12$. Therefore, it seems reasonable to
assume that above this doping level $x$ a change in the nature of the charge
carriers takes place. One could hypothesize that, as the formation of
bipolarons is stabilized with increasing density of the polaron gas \cite
{bipo}, bipolarons start playing a role in the optical absorption spectrum.
In a variety of other cuprates and manganates, the presence of bipolarons
has also been invoked to interpret a number of response-related properties 
\cite{alex}.

\section{Conclusions}

Starting from the many-polaron canonical transformations and the variational
many-polaron wave function (LDB) introduced in \cite{LDB} we have derived a
formula for the optical absorption coefficient $%
\mathop{\rm Re}%
[\sigma (\omega )]$\ of a many-polaron gas. We find that $%
\mathop{\rm Re}%
[\sigma (\omega )]$\ can be expressed in a closed analytical form in terms
of the dynamical structure factor $S(q,\omega )$\ of the electron (or hole)
system, equation (\ref{sig2d}) in 2D and (\ref{sig3d}) in 3D. In the present
approach, the electron-phonon coupling and the electron-electron many-body
effects formally decouple in the expression for $%
\mathop{\rm Re}%
[\sigma (\omega )]$. Therefore, the many-body effects in the electron (hole)
system can be taken into account by employing any desired approximation to
the dielectric response (Hartree-Fock, RPA, etc.) of this electron (hole)
system.

In the present work, the dynamical structure factor $S(q,\omega )$ of the
electron (or hole) gas was considered both in the Hartree-Fock and the RPA
approximation.

The main effect of the Pauli exclusion principle on the optical absorption
of the polaron gas turns out to be a shift of the oscillator strength
towards higher frequencies.\ This effect can be understood in terms of the
available initial and final states in the polaron-photon scattering process
and naturally invokes the Fermi energy $E_{\text{F}}$\ of the electron
(hole) gas.

The main effects in the case of the RPA approximation are an overall
reduction of the optical absorption at frequencies $\omega >\omega _{\text{LO%
}}$\ and the introduction of a novel absorption feature which we identified
as a plasmon-phonon peak. This plasmon-phonon peak shifts to higher
frequencies with increasing density, such that a double peak structure can
appear in the 3D many-polaron optical absorption spectrum, consistent with
the observed bimodal polaronic band in cadmium oxide \cite{Finkenrath}.

As a first application of the method presented here, we chose to investigate
the optical absorption of the interacting polaron gas in the RPA framework.
For Nd$_{2}$CuO$_{2-\delta }$ ($\delta <0.004$), similarities were observed
(see Fig. 4) between the line shape of the experimental d-band and the
many-polaron optical absorption as calculated here. To study the density
dependence, measurements (performed by Calvani and co-workers \cite{calva2}
for a family of Nd$_{2-x}$Ce$_{x}$CuO$_{4-y}$ materials) of the first
frequency moment of the optical absorption were compared to the results of
the present theory. We find a fair agreement for the samples with the lowest
densities (cerium doping $x<0.12$). A softening of the first frequency
moment of the optical absorption for the samples with higher densities\
(cerium doping $x>0.12$) is consistent with a change in the nature of the
charge carriers at a doping content $x=0.12$ inferred in ref. \cite{calva1}
from infrared absorption experiments.

\section*{Acknowledgments}

The authors like to acknowledge S. N. Klimin and V. M. Fomin for helpful
discussions and intensive interactions. We are indebted to P. Calvani for
fruitful discussions and for communication of experimental data. We thank F.
Brosens and L. F. Lemmens for discussions. One of us, J.T., (``Postdoctoraal
Onderzoeker van het Fonds voor Wetenschappelijk Onderzoek -- Vlaanderen''),
is supported financially by the Fonds voor Wetenschappelijk Onderzoek --
Vlaanderen (Fund for Scientific Research -- Flanders). Part of this work is
performed in the framework of the ``Interuniversity Poles of Attraction
Program -- Belgian State, Prime Minister's Office -- Federal Office for
Scientific, Technical and Cultural Affairs'' (``Interuniversitaire
Attractiepolen -- Belgische Staat, Diensten van de Eerste Minister --
Wetenschappelijke, Technische en Culturele Aangelegenheden''), and in the
framework of the FWO projects 1.5.545.98, G.0287.95, 9.0193.97, WO.025.99N
and WO.073.94N (Wetenschappelijke Onderzoeksgemeenschap, Scientific Research
Community of the FWO on ``Low Dimensional Systems''), and in the framework
of the BOF NOI 1997 and GOA\ BOF UA 2000 projects of the Universiteit
Antwerpen.

\newpage

\section*{Table}

{\bf Table I}: Material parameters used in the various figures. The physical
parameters for GaAs correspond to those of the GaAs-AlGaAs heterostructure 
\cite{wu2}; the material parameters for ZnO are taken from \cite{GreenBook}.
The physical parameters for the neodymium-cerium cuprate are taken from \cite
{calva1} and \cite{AlonsoPRB55}. ``n.a.'' (``not applicable'') means that
not enough data are available to estimate this material parameter.

\[
\begin{tabular}{lllll}
\hline\hline
{\bf material parameters:} &  & GaAs & ZnO & Nd$_{1.85}$Ce$_{0.15}$CuO$_{2}$
\\ \hline
phonon frequency & $\hbar \omega _{\text{LO}}=$ & $36.77$ meV & $73.27$ meV
& $74$ meV \\ 
dielectric constants & $\varepsilon _{0}=$ & $12.83$ & $8.15$ & n.a. \\ 
& $\varepsilon _{\infty }=$ & $10.9$ & $4.00$ & ca. $3.$ \\ 
band mass & $m_{\text{b}}=$ & $0.0657$ $m_{\text{e}}$ & $0.24$ $m_{\text{e}}$
& n.a. \\ 
coupling constant & $\alpha =$ & $0.068$ & $0.849$ & n.a. \\ 
polaron length unit & $a_{\text{HO}}=$ & $5.616$ nm & $2.082$ nm & n.a. \\ 
bohr radius & $a_{\text{B}}=$ & $8.7797$ nm & $0.882$ nm & n.a. \\ 
\hline\hline
\end{tabular}
\]

\bigskip \bigskip

\section{Figure captions}

{\bf Figure 1}\ : The real part of the optical conductivity (proportional to
the optical absorption coefficient) of an interacting large-polaron gas is
shown as a function of frequency, for a two dimensional gas (GaAs) from
equation \ref{sig2d}. The material parameters are given in Table I. The
dashed curve represents the one-polaron result, the full curve labeled
`Hartree-Fock structure factor' shows the result using the Hartree-Fock
approximation to the dynamical structure factor of the electron (hole)
system, and the full curve labeled `RPA structure factor' is the result in
the Random Phase approximation. The dotted vertical line indicates the
threshold frequency above which all polarons can be scattered into
unoccupied final states and participate in the absorption process. The broad
gray peak in the RPA curve is the plasmon-phonon contribution (see also
figure 2).

{\bf Figure 2} : The real part of the optical conductivity is shown as a
function of frequency for an interacting large-polaron gas in the 2D case
(left panel) and the 3D case\ (right panel). The material parameters used
here are given in Table I. The dashed curves represent the single polaron
spectra. The full curve represents the many-polaron spectrum. In this
figure, the plasmon-phonon contribution to the optical many-polaron spectrum
is shown as a shaded area. This contribution arises from a process where a
polaron, with the absorption of a photon, emits a phonon and a plasmon. The
inset shows the regions in the $q$-$v$ plane where the dynamical structure
factor $S(q,v)$ of the electron (or hole) system used in the optical
absorption formulas (\ref{sig3d},\ref{sig2d}) differs from zero; the Landau
damping region and the undamped plasmon branch can be distinguished.

{\bf Figure 3} : The real part of the optical conductivity is shown as a
function of frequency for different densities of an interacting
large-polaron gas, in the 2D case (left panel) and the 3D case\ (right
panel). The material parameters used for this figure are given in Table I.
For increasing density, the optical conductivity is reduced. Another effect
in the RPA approximation is the presence of a peak related to the undamped
plasmon branch (see figure 2) which shifts according to the plasma frequency.

{\bf Figure 4} : The infrared absorption of Nd$_{2}$CuO$_{2-\delta }$ ($%
\delta <0.004$) is shown as a function of frequency, up to 3000 cm$^{-1}$.
The experimental results of Calvani and co-workers \cite{calva2} is
represented by the thin black curve and by the shaded area. The so-called
`d-band' rises in intensity around 600 cm$^{-1}$ and increases in intensity
up to a maximum around 1000 cm$^{-1}$. The dotted curve shows the single
polaron result. The full black curve represents the theoretical results
obtained in the present work for the interacting many-polaron gas with $%
n=1.5 $ $10^{17}$ cm$^{-3}$, $\alpha =2.1$ and $m_{\text{b}}=0.5$ $m_{\text{e%
}}$.

{\bf Figure 5}\ : The normalized first frequency moment of the optical
absorption spectra is shown as a function of the density (expressed through
the Fermi wave vector). The squares represent the experimental results of
Calvani and co-workers \cite{calva2} in a family of Nd$_{2-x}$Ce$_{x}$CuO$%
_{4}$ materials. The dashed curve shows the results from the theoretical
two-dimensional many-polaron optical absorption, obtained by integrating all
frequencies in the calculation of the first frequency moment. The full curve
shows the theoretical results obtained by integrating up to a cut-off
frequency, which is chosen at $10000$ cm$^{-1}$ and which corresponds to the
maximum frequency in the experiment \cite{calva2}. The material parameters
are listed in the figure, and the effect of choosing a different electron
band mass is illustrated in the inset. The points with $x<0.12$, to the left
of the vertical dotted line, show agreement with the theoretical result from
the many-polaron theoretical optical absorption, but it is clear that the
experimental cutoff frequency has to be taken into account. For the samples
with $x>0.12$, a discrepancy between the theoretically predicted first
frequency moment and the observed first frequency moment is consistent with
a possible insulator-to-metal transition at $x=0.12$ \cite{calva1}.

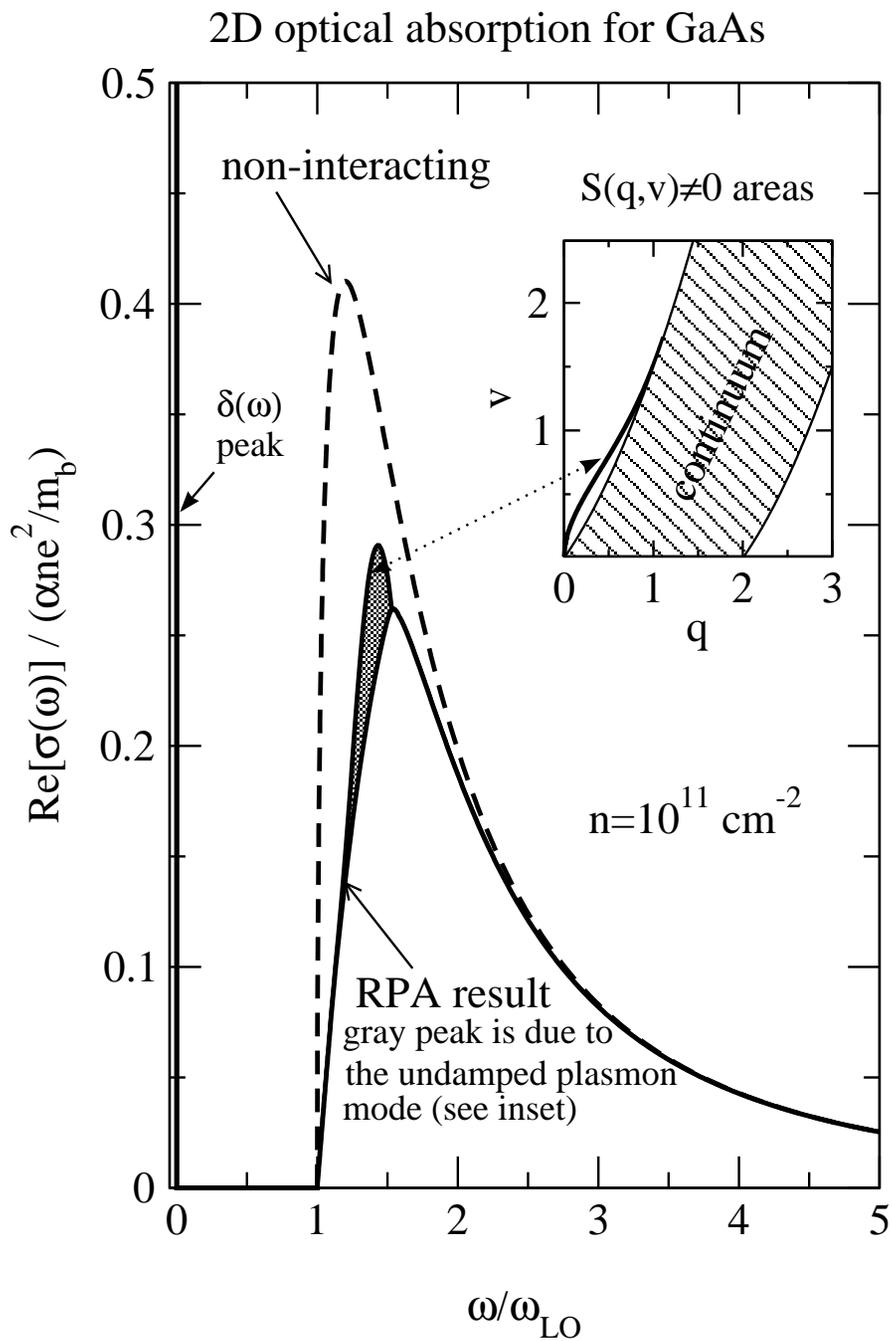
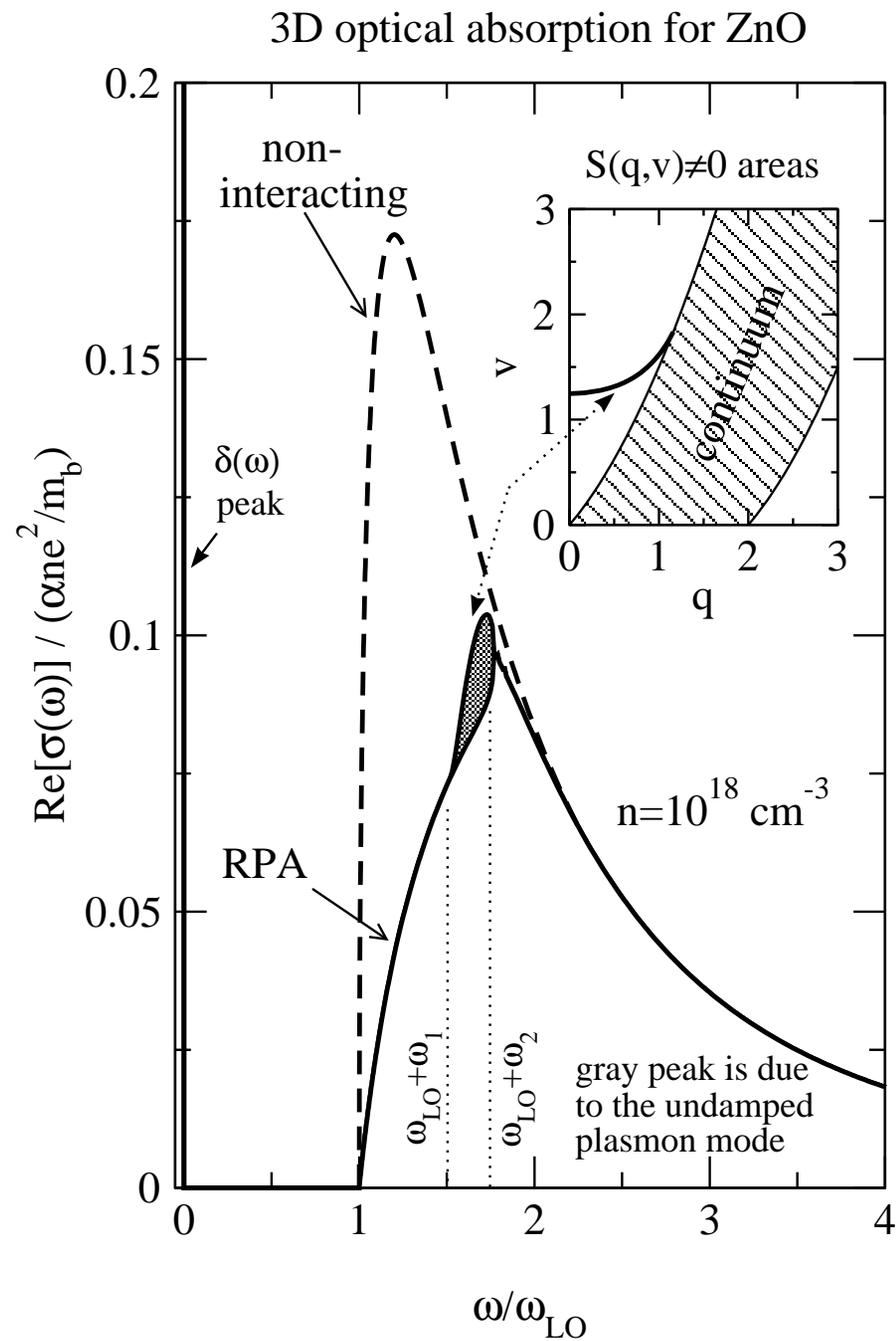
\end{document}